\documentclass[pra,twocolumn,showpacs,superscriptaddress,floatfix, nofootinbib]{revtex4-2}
\usepackage[caption=false]{subfig}
\usepackage{amsmath,graphicx}
\usepackage{amsmath,amssymb,mathrsfs,esint}
\usepackage{multirow}
\usepackage{algorithm}
\usepackage[noend]{algpseudocode}
\usepackage{comment}
\usepackage{tabularx}
\usepackage{bm}
\usepackage[normalem]{ulem}
\usepackage[bookmarks=true,
   colorlinks=true,
   linkcolor=blue,
   urlcolor=blue,
   citecolor=blue,
   bookmarks=true,
   hyperindex=true
]{hyperref}
\usepackage{svg}
\usepackage{xifthen}
\usepackage{listings}

\definecolor{bgc}{rgb}{0.95,0.95,0.94}
\definecolor{com}{rgb}{0,0.6,0}
\lstdefinestyle{code_listing}{
    basicstyle=\small\ttfamily,
    keywordstyle=\bfseries,
    backgroundcolor=\color{bgc},
    commentstyle=\color{com},
    columns=fixed,
    frame=trbl,
    framesep=1pt,
}
\newcommand{\bra}[1]{\ensuremath{ \langle #1| }}
\newcommand{\ket}[1]{\ensuremath{ |#1\rangle }}
\newcommand{\anc}{\ensuremath{ \sf ancillary }}
\newcommand{\gate}[4]{\ensuremath{
    {\sf #1}_{#3}^{#4} \ifthenelse{\isempty{#2}}{}{\left(#2\right)}
}}
\newcommand{\ph}[3][]{\gate{Ph}{#1}{#2}{#3}}
\newcommand{\rz}[3][]{\gate{RZ}{#1}{#2}{#3}}
\newcommand{\rp}[3][]{\gate{R}{#1}{#2}{#3}}
\newcommand{\xx}[3][]{\gate{XX}{#1}{#2}{#3}}

\newcommand{\cz}[1]{\gate{CZ}{}{#1}{}}
\newcommand{\mcz}[1]{\gate{MCZ}{}{#1}{}}
\newcommand{\cx}[1]{\gate{CX}{}{#1}{}}
\newcommand{\mcx}[1]{\gate{MCX}{}{#1}{}}

\newcommand{\id}[1][]{\gate{Id}{}{#1}{}}

\newcommand{\On}[1]{\ensuremath{ O\left({#1}\right) }}
\newcommand{\mi}{\mathrm{i}}
\newcommand{\pauli}[2]{\ensuremath{
    \sigma_{#1}^{#2}
}}
\newcommand{\one}[1]{\ensuremath{
    \varepsilon^{#1}
}}
\newcommand{\One}[1]{\gate{E}{}{}{#1}}

\graphicspath{ {./images/} }

\begin{document}

\title{Transpiling Quantum Assembly Language Circuits to a Qudit Form}

\author{Denis A. Drozhzhin}
\affiliation{National University of Science and Technology ``MISIS”,  Moscow 119049, Russia}
\author{Anastasiia S. Nikolaeva}
\affiliation{National University of Science and Technology ``MISIS”,  Moscow 119049, Russia}
\author{Evgeniy O. Kiktenko}
\affiliation{National University of Science and Technology ``MISIS”,  Moscow 119049, Russia}
\author{Aleksey K. Fedorov}
\affiliation{National University of Science and Technology ``MISIS”,  Moscow 119049, Russia}

\begin{abstract}
In this paper, we introduce the workflow for converting qubit circuits represented by Open Quantum Assembly format (OpenQASM, also known as QASM) into the qudit form for execution on qudit hardware and provide a method for translating qudit experiment results back into qubit results.
We present the comparison of several qudit transpilation regimes, which differ in decomposition of multicontrolled gates: \textbf{qubit} as ordinary qubit transpilation and execution, \textbf{qutrit} with $d{=}3$ levels and single qubit in qudit, and \textbf{ququart} with $d{=}4$ levels and 2 qubits per ququart.
We provide several examples of transpiling circuits for trapped ion qudit processors, which demonstrate potential advantages of qudits.
\end{abstract}

\maketitle

\section{Introduction}

A digital model of quantum computing relies on performing quantum logical operations under qubits, which are quantum analogs of classical bits allowing superposition states~\cite{Brassard1998,Ladd2010,Fedorov2022}.
It is expected that quantum processors of sufficiently high performance may be superior to classical counterparts in various computational problems~\cite{Fedorov2022,Gambetta2024},
such as prime factorization~\cite{Shor1994}, optimization~\cite{Blekos2023}, and simulating complex (quantum) systems~\cite{Lloyd1996,Daley2022}.
Key building blocks for realizing quantum processors based on various physical platforms,
such as superconducting circuits~\cite{Martinis2019,Pan2021}, semiconductor quantum dots~\cite{Loss1998,Vandersypen2022,Morello2022,Tarucha2022}, photonic systems~\cite{Pan2020,Lavoie2022},
neutral atoms~\cite{Lukin2021,Browaeys2021,Browaeys2020-2,Saffman2022},
and trapped ions~\cite{Monroe2017,Blatt2012,Blatt2018}
have been demonstrated.
Although quantum advantage has been shown in several experiments with noisy intermediate-scale quantum (NISQ) devices~\cite{Martinis2019,Pan2020,Pan2021},
finding the path towards large-scale quantum computing remains an open question.

A promising approach to scaling ion-based quantum processors is to use additional levels for encoding quantum information, which is at the heart of the concept of qudit-based quantum processors.
Qudit-based quantum information processing has been widely studied both theoretically and experimentally during the last decades~\cite{Farhi1998,Kessel2002,Muthukrishnan2000,Nielsen2002,Berry2002,Klimov2003,Bagan2003,Vlasov2003,Clark2004,Leary2006,Straupe2006,Straupe20062,Ralph2007,White2008,Straupe2008,Ionicioiu2009,Ivanov2012,
Kiktenko2015,Kiktenko2015-2,Song2016,Bocharov2017,Gokhale2019,Pan2019,Low2020,Martinis2009,White2009,Straupe2010,Wallraff2012,Mischuck2012,Gustavsson2015,Martinis2014,Ustinov2015,Morandotti2017,Balestro2017,Low2020,Sawant2020,Senko2020,Pavlidis2021,Rambow2021,OBrien2022,Nikolaeva2022,Nikolaeva2023}.
Experimental results include demonstrations of qudit processors based on trapped ions~\cite{Ringbauer2021,Aksenov2023,Ringbauer20232,Zalivako2024}, superconducting circuits~\cite{Hill2021,Schuster2022}, and quantum light~\cite{OBrien2022}.
Specifically, in the case of trapped ions, high-fidelity control over multilevel systems has been shown~\cite{Ringbauer2021,Aksenov2023,Ringbauer20232,Low2023,Zalivako2024}.
Qudits can be used both for storing multiple qubits and for using higher levels as ancillas in the case of the decomposition of multiqubit gates.
However, as soon as the majority of algorithms are formulated in the qubit form, in order to use qudits, one needs a procedure for transforming qubit circuits in the qudit form to achieve an advantage in terms of the resulting fidelity.

In this work, we focus on the problem of efficient transforming quantum circuits that are expressed via Quantum Assembly (QASM) format to the qudit form.
We focus our attention on this format since QASM is the widely used representation of qubit circuits, and many quantum frameworks can handle QASM code according to the desired algorithm.
Frameworks \textit{qiskit}~\cite{Qiskit} and \textit{cirq}~\cite{Cirq} are commonly used to implement qubit algorithms and run them on quantum hardware, with the ability to optimize qubit circuits and perform topology-aware multiqubit transpilation.
Processing of qudit circuits is also the topic of several research, such as the simulation platform for hybrid quantum systems \textit{QuDiet}~\cite{Chatterjee2023}, which introduces a special QASM format for qudit execution, or the numerical qudit optimization framework \textit{bqskit}~\cite{Bqskit}.
The distinguishing feature of our transpilation approach is to take qubit QASM circuit, store qubit into qudits, execute circuit on qudit device, and provide qubit results back to a user.
From this perspective, our technique provides a seamless qubit circuit interface to the user and utilizes the benefits of qudits with different dimensionality.

To maximize the potential of qudits to execute qubit circuits, we introduce several key concepts that cannot be reached using earlier developed quantum circuit frameworks.
Firstly, in our paper, we introduce the idea of a customizable transpiling process with target device description using runtime.
While other frameworks often operate with native gates as unitary matrices to perform numerical optimization \cite{Bqskit}, our transpiler omits a unitary representation of native gates and relies on analytical decompositions and composition rules in terms of native gates and mathematical expressions.
These rules can be defined for any quantum device to use in our transpiler seamlessly.
Moreover, we propose to use for these rules the similar QASM description format (analytical formulas in the form of syntactic rules).
These facts simplify the process of operating with the transpiler
and greatly reduce the computing complexity of transpiling and optimizing qubit circuits.

The second idea is the process of unmapping samples obtained via the qudit quantum computer.
Using the qubit-to-qudit mapping from transpiler, one can convert these results back to qubit form.
This process provides the opportunity to use qudit quantum hardware or a simulator to run any qubit circuit in QASM form and obtain the circuit’s statistics corresponding to qubits defined in the QASM file.

The third is the idea of efficient multiqubit gate decomposition in terms of qudit gates.
Using qudit-based transpilation techniques, we expect to benefit in quantum circuit fidelity due to reducing the number of noisy two-qudit operations.
We present a comparison of several qudit execution regimes, whose main difference is in the decomposition of multicontrolled gates: \textbf{{qubit}} as ordinary qubit transpilation and execution, \textbf{qutrit} with $d{=}3$ levels and single qubit in qudit, and \textbf{ququart} with $d{=}4$ levels and 2 qubits per ququart.

The paper is organized as follows.
In Section~\ref{qasm}, we briefly revise the specification of the QASM format.
In Section~\ref{qudit-circuit}, we discuss basic gates for qudit circuit construction.
In Section~\ref{ion-qc-description}, we consider the trapped ion qudit quantum processor as a concrete example of a hardware platform for the transpilation process and provide a description of JSON format for the record of the qudit circuit.
In Section~\ref{transpiler}, we present a developed general transpilation workflow and a concrete realization of transpilation methods for input QASM circuits, which allows one to obtain implementable on a hardware qudit circuit for ion qudits with $d{=}3$ and $d{=}4$ levels.
In Section~\ref{comparison}, we benchmark a transpiler developed on widely used quantum algorithms and compare its different regimes with a qubit qiskit transpiler and between each other.
Finally, we conclude this paper in  Section~\ref{conclusion}.

\section{Quantum Circuits in the QASM Format}\label{qasm}

The QASM format~\cite{OpenQASM} is a widely used format for writing qubit quantum circuits in gate-based model in a text form.
QASM format could be generated using one of the quantum computing frameworks (e.g., qiskit~\cite{Qiskit}), from user-defined logic or could be written manually.
It is allowed due to the syntactical and semantic simplicity of the QASM representation, which consists of gates, measures, and barrier operations along with their conditional variants.
This set of operations is sufficient for quantum program execution on real hardware or on a simulator.

QASM was designed to formalize quantum computations and fully describe quantum circuits.
A QASM file is composed of a series of instructions, each of which represents a specific operation that is to be executed on the hardware (see Appendix~\ref{app:qasm-format}).

QASM format does not specify what type of quantum system it uses: qubit or qudit.
Although all qubit operations can be expressed using QASM, certain qudit operations may not be expressible due to the lack of a suitable description for the level structure and operations associated with qudits.
So, in our experiments, we use the QASM format for qubit quantum circuits, which describe the quantum logic of a program, and as input to the transpiler, which converts QASM circuit into qudit form for further execution on qudit hardware.
We also provide JSON format specification for qudit circuit description in Section~\ref{ion-qc-description}.

\section{Qudit Circuits}\label{qudit-circuit}

By analogy with qubit-based quantum computing, one can implement quantum circuits with $d$-level quantum systems, qudits.
On the existing qudit-based hardware, single-qudit operation \gate{U}{}{d}{ij}, which acts on a $d$-level qudit, is usually implemented as a unitary $2{\times}2$ matrix acting on a linear span of $i$-th and $j$-th levels (see an example for a trapped ion platform~\cite{Ringbauer2021,Aksenov2023} and for a superconducting platform~\cite{Schuster2022}).
Single-qudit operation \gate{U}{}{d}{ij} can be defined with the use of a qudit extension of Pauli matrices \pauli{x}{ij}, \pauli{y}{ij} and \pauli{z}{ij}, acting on $i$-th and $j$-th levels, that are the analogs to the qubit Pauli matrices extended {with zeroes:}
\begin{gather}
    \begin{aligned}
        \pauli{x}{ij} &= \ket{j}\bra{i} + \ket{i}\bra{j},
        \\
        \pauli{y}{ij} &= \mi\ket{j}\bra{i} - \mi\ket{i}\bra{j},
        \\
        \pauli{z}{ij} &= \ket{i}\bra{i} - \ket{j}\bra{j},
    \end{aligned}
\end{gather}
where $i,j\in\{0,\dots,d-1\}$; $d$ is the number of levels in the qudit; and $\mi$ stands for an imaginary unit.
These matrices satisfy the following relations:
\begin{gather}
    \left.\pauli{a}{ij}\pauli{b}{ij}\right|_{a,b\in\{x,y,z\}} = \begin{cases} \one{ij} & \text{if $a=b$,} \\ -\pauli{b}{ij}\pauli{a}{ij} & \text{otherwise,} \end{cases}
\end{gather}
where $\one{ij} = \ket{i}\bra{i} + \ket{j}\bra{j}$ is the identity matrix acting on $i$-th and $j$-th levels.

Single-qudit native operations for a trapped ion and a superconducting platform are defined as a rotation with two angle parameters $\theta$ and $\phi$.
Parameter $\theta$ specifies the angle of rotation that is determined by pulse length in a qudit system.
Parameter $\phi$ specifies the axis of rotation in the $XY$ plane within the Bloch sphere.
Using a qudit extension of Pauli matrices, we can obtain a matrix representation for a single-qudit operation \rp{}{ij}:
\begin{gather}
    \rp[\theta, \phi]{}{ij}=\exp\left(-\mi\theta\pauli{\phi}{ij}\right),
\end{gather}
where
\begin{equation}
     \pauli{\phi}{ij} = \cos\left(\phi\right)\pauli{x}{ij}+\sin\left(\phi\right)\pauli{y}{ij},
\end{equation}
and the following symmetry relations are fulfilled:
\begin{align}
    {\rp[\theta, \phi]{}{ij}}^\dagger &= \rp[-\theta, \phi]{}{ij}, \\
    \rp[\theta, \phi]{}{ji} &= \rp[\theta, -\phi]{}{ij}, \\
    \rp[-\theta, \phi]{}{ij} &= \rp[\theta, \phi \pm \pi]{}{ij}, \\
    \rp[\theta + 2\pi n, \phi]{}{ij} &= \rp[\theta, \phi]{}{ij}, \\
    \rp[\theta, \phi + 2\pi m]{}{ij} &= \rp[\theta, \phi]{}{ij},
\end{align}
where the arbitrary integers are $n$ and $m$.

We note that we define \rp{}{ij} gate without $\frac{1}{2}$ factor to preserve $2\pi$ periodicity for all parameters.
However, typical definitions of single-qubit gate \rp{\sf qb}{} and single-qudit gate \rp{\sf qd}{ij} depend on the $\frac{\theta}{2}$ parameter:
\begin{gather}
\rp[\theta, \phi]{\sf qb}{} = \exp\left( -\mi\frac{\theta}{2}\pauli{x}{} \right),
\\
\rp[\theta, \phi]{\sf qd}{ij} = \exp\left(-\mi\frac{\theta}{2}\pauli{\phi}{ij}\right).
\end{gather}
While the qubit operation satisfies the relation $\rp[\theta + 2\pi, \phi]{\sf qb}{}=-\rp[\theta, \phi]{\sf qb}{}$ and, therefore, two operations become equivalent up to global phase factor $-1$, the
qudit \rp{\sf qd}{ij} operation does not hold this equivalence:
\begin{multline}
\rp[\theta + 2\pi, \phi]{\sf qd}{ij} =\\= \One{ij} - \one{ij} \cos\left(\frac{\theta}{2}\right) + \mi \pauli{\phi}{ij} \sin\left(\frac{\theta}{2}\right),
\end{multline},
\begin{equation}
\rp[\theta + 2\pi, \phi]{\sf qd}{ij} + \rp[\theta, \phi]{\sf qd}{ij} = 2 \One{ij},
\end{equation}
where
\begin{equation}
    \One{ij} = \sum_{k\ne i,j} \ket{k}\bra{k} = \id{} - \one{ij}
\end{equation}
and $\id{}$ is an identity matrix.

Along with \rp{}{ij} operation, qudit devices provide the possibility to apply a phase operation \ph{}{i} and $Z$ rotation \rz{}{ij}, which can be often implemented virtually (see~\cite{Zalivako2024} for the trapped ions and~\cite{Schuster2022} for the superconducting platform), and hence, does not contribute to generating errors:
\begin{gather}
    \ph[\theta]{}{i} = \exp\left(\mi\theta\one{i}\right) = \One{i} + e^{\mi\theta} \one{i}, \\
    \rz[\theta]{}{ij} = \exp\left(-\mi\theta\pauli{z}{ij}\right) = \One{ij} + e^{-\mi\theta} \one{i} + e^{\mi\theta} \one{j}.
\end{gather}

Another crucial type of operation for quantum computations is entangling a two-qudit gate.
The specific form of two-qudit operations varies depending on the physical platform considered.
In theoretical papers, two-qudit ${\sf CZ}^{i|j}$ and ${\sf C X}^{i|jk}$ gates
are commonly considered:
 \begin{equation}
    {\sf CZ}^{i|j}:
    \left\{\begin{tabular}{l l l l}
        &$\ket{i,j}$ & $\mapsto -\ket{i,j}$, &\\
	&$\ket{x,y}$ & $\mapsto \ket{x,y}$ & \quad if $x\neq i\text{~or~}y\neq{j}$,
    \end{tabular}\right.
\end{equation}
\begin{equation}
    {\sf CX}^{i|jk}:
    \left\{\begin{tabular}{l l l l}
        &$\ket{i,j}$ & $\mapsto \ket{i,k}$, &\\
	&$\ket{i,k}$ & $\mapsto \ket{i,j}$, &\\
	&$\ket{x,y}$ & $\mapsto \ket{x,y}$ & \quad if $x\neq i\text{~or~} y\neq{j,k}$.
    \end{tabular}\right.
\end{equation}

However, trapped ion qudit-based quantum computers operate with a M\o{}lmer--S\o{}rensen gate~\cite{Molmer-Sorensen1999-2} \xx{}{ij|kl}:
\begin{equation} \label{eq:MS}
    \xx[\theta]{}{ij|kl} = \exp\left(-\mi\theta\pauli{x}{ij}\otimes\pauli{x}{kl}\right),
\end{equation}
where the pairs of levels $i,j$ and $k,l$ refer to the first and second qubits, respectively.
Within superconducting processors, an iSWAP gate of the following form can be typically implemented:
\begin{equation} \label{eq:iSWAP}
    {\sf iSWAP}^{ij|kl}(\theta):
    \begin{cases}
        \ket{i,k} \mapsto {e^{\imath\theta}}\ket{j,l}, \\
        \ket{j,l} \mapsto {e^{\imath\theta}}\ket{i,k},
    \end{cases}.
\end{equation}

Notably, a real quantum device implies selection rules, which determine allowed level pairs for single-qudit and two-qudit gates.
We define a \gate{SWAP}{}{}{ij} operation between levels $i$ and $j$ as \rp[\frac{\pi}{2}, \frac{\pi}{2}]{}{ij} to solve the issue.
Effectively, it swaps the population between $i$-th and $j$-th levels, allowing us to decompose any \rp{}{ij} into the sequence of allowed swaps and transitions.
For any $i$, $j$ and $s$, the following identity holds:
\begin{equation} \label{eq:swap-levels}
    \gate{G}{}{n}{i}=\gate{SWAP}{}{n}{is}\circ\gate{G}{}{n}{s}\circ\gate{SWAP}{}{n}{si},
\end{equation}
where \gate{G}{}{n}{j} is the pattern for operation that involves the $i$-th level in $n$-th qudit in register, e.g., \rp{n}{i\cdot}, \xx{n\cdot}{i\cdot\cdot\cdot}, \xx{\cdot n}{\cdot\cdot i\cdot}, etc., and $\circ$ denotes the operations' composition, where operations are applied in a \textit{left-to-right} order.
Also, for this decomposition to be valid, the operation pattern \gate{G}{}{n}{j} cannot involve an $s$-th level.
For example, for \xx{}{} gate, the level swap is performed as follows:
\begin{multline} \label{eq:swap-levels:xx}
    \xx[\theta]{n|m}{ij|kl}=\\=\left(\gate{SWAP}{}{n}{is}\otimes\id[m]\right)\circ\xx[\theta]{n|m}{sj|kl}\circ\left(\gate{SWAP}{}{n}{si}\otimes\id[m]\right).
\end{multline}

\section{Trapped Ion Qudit Quantum Computer}\label{ion-qc-description}

As a concrete example of a qudit device for the transpilation process, we consider the trapped ion qudit-based processor developed in Refs.~\cite{Aksenov2023,Zalivako2024}.
The authors implemented qudit-trapped ion computations (in this work, we consider at most $d{=}4$) using the \rp{\text{ion}}{0i}, \ph{\text{ion}}{i} and \xx{\text{ion}}{01|01} native gate set.
These gates are equivalent to the defined ones in the previous section with parameter modifications:
\begin{equation} \label{eq:ion}
\begin{aligned}
    \ph[\theta]{\text{ion}}{i} &= \ph[\theta]{}{i}, \\
    \rp[\theta, \phi]{\text{ion}}{0i} &= \rp[\frac{\theta}{2}, \phi]{}{0i}, \\
    \xx[\theta]{\text{ion}}{01|01} &= \xx[\theta]{}{01|01}.
\end{aligned}
\end{equation}

Considering the allowed level transitions for $^{171}\text{Yb}^{+}$ ion qudits and Equation~(\ref{eq:swap-levels}), it is possible to decompose any sequence of qudit operations \ph{}{i}, \rp{}{ij} and \xx{}{ij|kl} into the sequence of ion native gates \ph{\text{ion}}{i}, \rp{\text{ion}}{0i} and \xx{\text{ion}}{01|01} \cite{nikolaeva2023universal}.

Operations \ph{\text{ion}}{i}, \rp{\text{ion}}{0i} and \xx{\text{ion}}{01|01} on allowed levels are stored in an \textit{ion quantum computer} description format (see Appendix~\ref{app:iqc-json-format}).

\section{Qasm to Qudit Form Transpilation}\label{transpiler}

In our model, qudit circuits can only be constructed with the native gate set for a given quantum device.
Native gates are the operations physically allowed on the device, and, usually, a set of native gates contains single-particle and two-particle operations.
On the other hand, QASM is a more general quantum circuit description format.
It can represent not only a sequence of physical gates but rather quantum logic for this circuit with, for example, \gate{U}{\theta,\phi,\lambda}{}{} and controlled \cx{} or multiqubit Toffoli operations.
Hence, in some cases, even qubit circuits in QASM format could not be executed on the device as is.

\begin{figure}[b]
\centering
\includegraphics[width=\columnwidth]{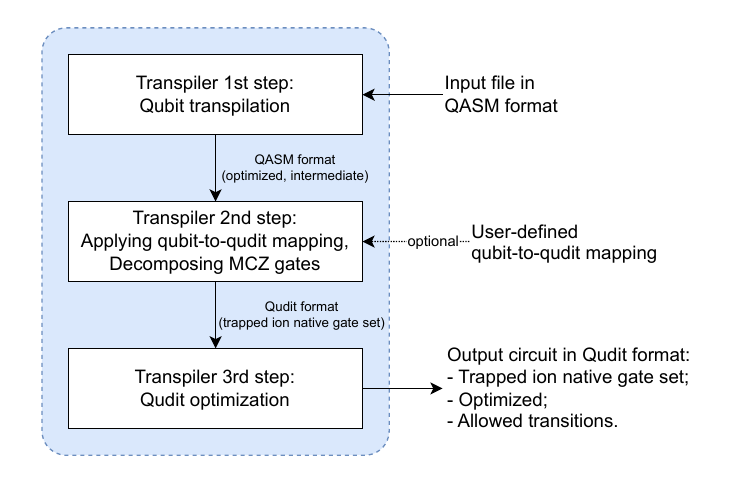}
\caption{General scheme of transpilation process and changing the format of the circuits (arrows with caption) during transpilation.}
\label{fig:process}
\end{figure}

The transpiler aims to bridge the gap between high-level software and hardware representations of a circuit and to facilitate the process of translating QASM code into a qudit circuit.
We divide the transpilation process into three steps, as depicted in Figure~\ref{fig:process}: qubit transpilation, qubit--qudit transpilation (or qudit transpilation, which is performed according to qubit-to-qudit mapping), and qudit circuit optimization.
At the end of transpilation, an optimized qudit circuit can be stored in a format suitable for execution on the target device.

In the first step, the 'pure' qubit transpilation  takes place.
It parses QASM code into an abstract syntax tree, which is used as a circuit representation within the transpiler, and converts a sequence of logical quantum gates into a native gate set using decompositions of the most used gates (e.g. \gate{U}{\theta, \phi, \lambda}{}{}, \gate{CX}{}{}{}, \gate{MCX}{}{}{}, etc.).
This process is divided into two phases: preprocessor and optimizer.
Each phase could be tuned according to the chosen target device.

At the first phase, the preprocessor performs parsing and checking the semantics of the QASM file.
Any encountered error in source code is supplemented with a descriptive error message and the line number and position to refer to it in source code.
Next, it expands all QASM gates into the native gate set for the device using the provided \textit{qelib1.inc} file, which consists of many applied gate declarations according to OpenQASM 2.0 standard.
After the preprocessor phase, the user obtains a sequence of native qubit gates.
Here, the optimizer could be enabled to reduce the number of operations and produce an equivalent gate sequence.

At the second phase, the optimizer undergoes the subsequences of gates from the source circuit and tries to match them with a predefined set of symmetry relations and gate contraction rules, which can be defined in a separate matcher file \textit{matcher.script}.
The matcher file represents the set of rules in the form of mathematical expressions that can be derived from target device operations' relations.
Due to such a form, this file can be written by developers and scientists for a given device without using a matrix representation of the native gate.

Along with the native gate set, the file with gate declarations \textit{qelib1.inc} and rule set \textit{matcher.script} represent the \textit{runtime} for the given target device.
We describe the content of \textit{qelib1.inc} and \textit{matcher.script} more precisely in Appendix~\ref{app:runtimes}.
For now, the transpiler supports one of the following runtimes and corresponding native gate sets:
\begin{itemize}
    \item \textit{Quantum emulator}:\\
    $\gate{U}{}{}{},\ \cx{}$.
    \item \textit{Trapped ion quantum computer}:\\
    $\rz{}{},\ \rp{}{},\ \xx{}{}$.
    \item \textit{Trapped ion intermediate representation}:\\
    $\rz{}{},\ \rp{}{},\ \cz{},\ \mcz{}$.
\end{itemize}
where $n$, $m$ and other lower indices denote qubits (or qudits) where the operation acts on.
\textit{Quantum emulator} and \textit{trapped ion quantum computer} runtimes are suitable for the case of qubit execution.
\textit{Trapped ion intermediate representation} is the special runtime, which should not resemble any quantum hardware or quantum simulator.
However, this runtime produces a qubit circuit with an extended gate set and passes it into the following step.
This extended circuit represents an intermediate format that allows the transpiler to choose a qudit decomposition for \rz{}{}, \rp{}{} and \cz{} for a given type of qudit.

Another point of introducing this runtime is that it contains decompositions for multiqubit gates (e.g., Fredkin \gate{CSWAP}{}{}{} or multicontrolled Toffoli \gate{MCX}{}{}{}) in terms of \mcz{} gates.
For many qudit parameters, this gate has special form in the qudit circuit that will arise during the second step of transpilation.
The resulting qudit circuit will only contain \xx{}{ij|kl} gates as two-qudit operations.
Additionally, the qudit optimization step could be enabled to reduce the length of the qudit sequence of gates.
Qudit optimization rules are described in the third step of the transpiler.

In the second step, the transpiler uses a qubit circuit transpiled with a \textit{trapped ion intermediate representation} runtime from the first step to produce a qudit circuit.
Along with the number of parameter of levels in qudit $d$, it uses a parameter $b$ that defines how many qubits from the original circuit can be embedded in each qudit from the generated qudit circuit.
Qudit, containing at least $2^b$ levels, is able to represent qubit register with $b$ qubits by placing register states on $2^b$ \textit{qubit} levels.
If the number of levels is $d>2^b$, then the remaining levels are considered as \textit{ancillary}.
An example of qudit structure is presented in Figure~\ref{fig:qudit-levels}.
Qubits ($d{=}2$) or qutrits ($d{=}3$) can only store a single qubit, whereas ququarts ($d{=}4$) can store one or two qubits.

\begin{figure}[t]
\centering
\begin{tabular}{c|c|c}
    qudit \quad & \ qubits $b{=}1$ \quad & \ qubits $b{=}2$ \\
    $\ket{0}_{\sf qd}$ & $\ket{0}_{\sf qb}$ & $\ket{0}_{\sf qb}\otimes\ket{0}_{\sf qb}$ \\
    $\ket{1}_{\sf qd}$ & $\ket{1}_{\sf qb}$ & $\ket{0}_{\sf qb}\otimes\ket{1}_{\sf qb}$ \\
    $\ket{2}_{\sf qd}$ & \anc & $\ket{1}_{\sf qb}\otimes\ket{0}_{\sf qb}$ \\
    $\ket{3}_{\sf qd}$ & \anc & $\ket{1}_{\sf qb}\otimes\ket{1}_{\sf qb}$ \\
    $\ket{4}_{\sf qd}$ & \anc & \anc \\
    \vdots & \vdots & \vdots
\end{tabular}
\caption{Qudit levels to qubit state mapping for $b{=}1$ and $b{=}2$.
For any qudit levels $d$ and qubits in qudit $b$, all qudit levels are divided into $2^b$ \textit{qubit} levels and $d{-}2^b$ \textit{ancillary} levels.
Each \textit{qubit} level digit is converted into qubit state using its binary representation, whereas \textit{ancillary} levels are used for specific decompositions.}
\label{fig:qudit-levels}
\end{figure}

\begin{figure}[t]
\centering
\includegraphics[width=\columnwidth]{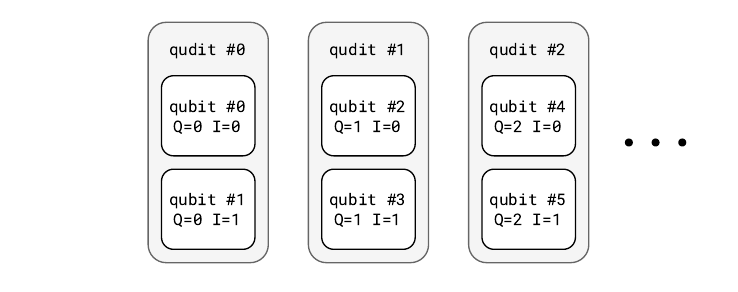}
\caption{Example for a straightforward mapping for $b{=}2$.
White blocks represent the initial qubits from the QASM circuit.
Qubits are uniquely assigned with $Q_n$ and $I_n$ indices.
The straightforward mapping assigns $Q=0$ to the first pair of qubits, $Q=1$ to the second pair, and so on.
Within a single pair, qubits are distinguished by index $I$.}
\label{fig:qb-qd-mapping}
\end{figure}

To store all qubits from the QASM circuit into qudits, the transpiler produces a qubit--qudit mapping.
The result is two sets of indices  $Q_n$ and $I_n$ for any qubit index ${n\in\{0, 1, \dots, N{-}1\}}$, where $N$ is the size of the register in the original circuit:
$Q_n$ stands for qudit index in the qudit register and ${I_n\in\{0, 1, \dots, b{-}1\}}$ stands for a qubit index within a single qudit.
The current implementation of the transpiler assigns indices in a straightforward manner, as shown in Figure~\ref{fig:qb-qd-mapping}: $Q_n{:=}\lfloor n\ /\ b\rfloor$ and $I_n{:=}\left(n\ \mathrm{mod}\ b\right)$ .
{Although the straightforward mapping is sufficient for qubit and qutrit transpilation, it may produce a less efficient ququart circuit.
In this case, users can specify custom mapping based on their circuit.
The aspect of proper mapping will be discussed in Section~\ref{comparison}.
Also, it is intended to develop an algorithm that uses heuristics to generate proper mapping that minimizes the number of two-qudit gates in the resulting circuit~\cite{Nikolaeva2021}.}

These indices are used to convert the qubit circuit into qudit operations.
Single-qubit gates for the case $b{=}1$ (implies $\forall n: I_n=0$) are mapped as
\begin{gather}
    \rp[\theta, \phi]{n}{}\rightarrow\rp[\theta, \phi]{Q_n}{01}, \\
    \rz[\theta]{n}{}\rightarrow\ph[\theta]{Q_n}{1}.
\end{gather}

For $b>1$, the transpiler decomposes \rp{}{} and \rz{}{}, considering the position of the qubit within a single qudit ($I_n$ index).
This decomposition consists of $2^{b-1}$ qudit gates \rp{}{ij} and \ph{}{k}, respectively.
Levels $i$, $j$ and $k$  are obtained explicitly from the $I_n$ index of a qubit.
For example, for $b{=}2$,
\begin{gather}
    \rp[\theta, \phi]{n}{}\rightarrow
    \begin{cases}
        \rp[\theta, \phi]{Q_n}{01}\circ\rp[\theta, \phi]{Q_n}{23} & \text{if $I_n=0$}, \\
        \rp[\theta, \phi]{Q_n}{02}\circ\rp[\theta, \phi]{Q_n}{13} & \text{if $I_n=1$},
    \end{cases} \\
    \rz[\theta]{n}{}\rightarrow
    \begin{cases}
        \ph[\theta]{Q_n}{1}\circ\ph[\theta]{Q_n}{3} & \text{if $I_n=0$}, \\
        \ph[\theta]{Q_n}{2}\circ\ph[\theta]{Q_n}{3} & \text{if $I_n=1$}.
    \end{cases}
\end{gather}

Notably, decompositions of multiqubit gates can benefit for some qudit parameters $d$ and $b$.
For this case, we used a special \textit{trapped ion intermediate representation} runtime in the first stage on qudit transpilation.
With this runtime, we obtained an optimized circuit that contains multiqubit \mcz{} gates, and in the second stage, we utilized these gates to produce the efficient decomposition using different values for transpilation parameters $d$ and $b$.
In our paper, we focus on several cases of qudit transpilation.

The first case is that of qutrits ($d{=}3$) or ququarts ($d{=}4$) with $b{=}1$, having a special \mcz{} gate decomposition~\cite{Nikolaeva2022},
which efficiently uses higher levels and produces a sequence of $2N-3\in\On{N}$ \xx{}{01|01} operations,
whereas a straightforward qubit algorithm~\cite{Barenco1995} will require \On{2^N} \cx{}/\cz{} operations or \On{N} operations with \On{N} ancillary qubits (here, $N$ is the number of qubits involved).

The second case is that of qudits with $b>1$ and $d{=}2^b$.
In this case, all levels are occupied by qubits, thus no ancillary levels can be used.
\mcz{} gates can be represented in terms of \cz{} gates using standard qubit decomposition.
However, some of the operations will end up acting on two qubits in a single qudit, and hence, will be decomposed into \ph{}{m} gate instead of \xx{}{ij|kl} for qubits in different qudits~\cite{Kiktenko2023}.
Considering the matrix representation of \cz{}, the following definition for a two-qudit gate is obtained:
\begin{gather}
    \cz{nm}\rightarrow
    \begin{cases}
        \ph[\pi]{Q_n}{3} & \text{if $Q_n=Q_m$} \\
        \xx[\pi]{Q_n Q_m}{i3|j3} & \text{if $Q_n\neq Q_m$}
    \end{cases}
\end{gather}
where $i=2^{I_n}$ and $j = 2^{I_m}$.

Due to the permutation symmetry of \mcz{} gate, we can virtually swap qubits to obtain an efficient circuit.
There is no issue in the case of qubit transpilation; however, for many qudit cases, it can minimize the amount of qudit \xx{}{} gates in template decomposition.
The analysis of qubit permutations in \mcz{} for qudits will be considered in future works.

The third step of transpilation consists of optimizing the qudit circuit according to the allowed level transitions and the production of the qudit circuit descriptor for execution, as described in Section~\ref{ion-qc-description}.
The circuit after the second step consists of qudit operations \ph{}{i}, \rp{}{ij} and \xx{}{ij|kl}, while the trapped ion computer accepts operations with specific transition levels only.
The transpiler uses level swap operations \gate{SWAP}{}{}{ij} from Equation~(\ref{eq:swap-levels}) whenever possible and keeps only the operations allowed by a selection rule:
\begin{gather}
    \gate{G}{}{n}{i}\rightarrow\gate{SWAP}{}{n}{is}\circ\gate{G}{}{n}{s}\circ\gate{SWAP}{}{n}{si}, \\
    \gate{G}{}{n}{i}\in\{\rp{n}{i\cdot(\cdot i)}, \xx{n\cdot}{i\cdot|\cdot\cdot(\cdot i|\cdot\cdot)}, \xx{\cdot|n}{\cdot\cdot|i\cdot(\cdot\cdot|\cdot i)}\},
\end{gather}
where \gate{G}{}{n}{i} is a pattern for operation acting at the $i$-th level in  the $n$-th qudit.
Using this rule, the transpiler is able to convert any \rp{}{ij} or \xx{}{ij|kl} operation into the sequence of allowed transitions for a given adjacency graph between levels in a given qudit device.

Along with the conversion of the allowed operations, the transpiler optimizes the resulting qudit circuit.
Although most initial gates from QASM are already optimized during qubit transpilation, the circuit has the potential for optimization after applying level swaps and \mcz{} gate decompositions.
This optimization step differs from qubit optimization due to more specific matching rules.
They include combinations of rotations and removing operations that are non-affecting on measurement results $\ph[\theta]{}{i}$:
\begin{flalign}
    \ph[\theta_1]{}{i}\circ\ph[\theta_2]{}{i} &\rightarrow \ph[\theta_1+\theta_2]{}{i} \\
    \ph[2n\pi]{}{i} &\rightarrow \id \\
    \rp[\theta_1,\phi]{}{ij}\circ\rp[\theta_2,\phi]{}{ij} &\rightarrow \rp[\theta_1+\theta_2,\phi]{}{ij} \\
    \rp[\theta, \phi]{}{ji} &\rightarrow \rp[\theta,-\phi]{}{ij} \\
    \rp[2n\pi, \phi]{}{ij} &\rightarrow \id \\
    \xx[\theta_1]{}{ij|kl}\circ\xx[\theta_2]{}{ij|kl} &\rightarrow \xx[\theta_1+\theta_2]{}{ij|kl} \\
    \xx[2n\pi]{}{ij|kl} &\rightarrow \id
\end{flalign}
Rules are designed to preserve the unitary matrix of the circuit, to reduce the amount of
\rp{}{ij} and \xx{}{ij|kl} operations and to try to move \ph{}{i} to the end of the circuit.
The last is used to remove trailing phases since they are not affecting measuring in computational $Z$-basis.
Yet this step doesn't preserve unitary matrix of the circuit and is considered optional.

At the end, the transpiler is able to produce a qudit circuit in the format presented in Section~\ref{qudit-circuit}.
Since this format supports several circuits in  a single JSON file, all input QASM files are placed into a single output file.

Notably, qudit circuits and qudit computers operate with qudits and dits (classical numbers $0,1,\cdots,d{-}1$), rather than qubits and bits.
This fact leads to the problem of \textit{unmapping} qudit experiment samples into qubit results.
Given qudit computer execution output is an array of qudit states and samples counts, we only have to convert qudit states into qubit states using the mapping.

\begin{figure}[t]
\centering
\begin{tabular}{c|c|c}
    qudit \quad & \ strict \quad & \ non-strict \\
    $\ket{0}_{\sf qd}$ & $\ket{00}_{\sf qb}$ & $\ket{00}_{\sf qb}$ \\
    $\ket{1}_{\sf qd}$ & $\ket{01}_{\sf qb}$ & $\ket{01}_{\sf qb}$ \\
    $\ket{2}_{\sf qd}$ & $\ket{10}_{\sf qb}$ & $\ket{10}_{\sf qb}$ \\
    $\ket{3}_{\sf qd}$ & $\ket{11}_{\sf qb}$ & $\ket{11}_{\sf qb}$ \\
    $\ket{4}_{\sf qd}$ & \sf{excluded} & $\ket{11}_{\sf qb}$ \\
    $\ket{5}_{\sf qd}$ & \sf{excluded} & $\ket{11}_{\sf qb}$ \\
    \vdots & \vdots & \vdots
\end{tabular}
\caption{Two possible ways to convert qudit state back to qubit state.
The first is \textit{strict} mode which considers levels $\geq2^b$ as an error and excludes them from resulting samples.
The second is \textit{non-strict} mode, which converts these levels to the highest allowed level.}
\label{fig:unmapping}
\end{figure}

\begin{figure}[t]
\centering
\includegraphics[width=\columnwidth]{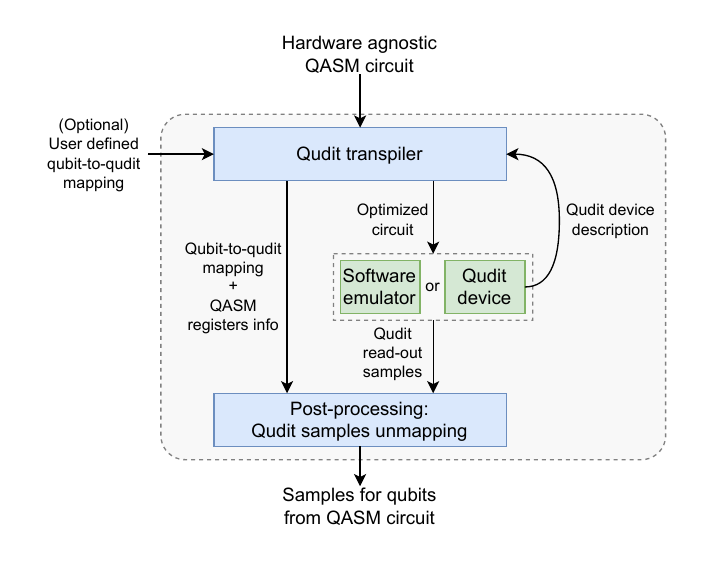}
\caption{Qudit transpilation, quantum execution and sample unmapping form the workflow for qubit circuit execution on qudit-based hardware.}
\label{fig:workflow}
\end{figure}

Qudit state is represented either as an array of numbers (e.g., \lstinline|[0, 1, 2, 1, 2]|), or as a string (e.g., \lstinline|"01212"|).
Firstly, since each qudit consists of $b$ qubits, dits $0,1,\cdots,2^{b}{-}1$ are the only values that should be observed in an experiment.
Although other dits with values $\geq2^b$ are still can be sampled, they are considered as an error and can be excluded from qubit samples (\textit{strict} mode) as shown in Figure~\ref{fig:unmapping}.
The other way to handle higher dits is to interpret them as the highest allowed level $2^{b}{-}1$ (\textit{non-strict} mode).
Secondly, dits can be converted into binary representation (\lstinline|0| $\rightarrow$ \lstinline|"000"|, \lstinline|1| $\rightarrow$ \lstinline|"001"|, etc.), where each bit corresponds to one of the qubit from the original circuit.
The index of the qubit can be obtained by inverting qubit-to-quqit mapping, using indices $Q_n$ and $I_n$.
For each dit at position $Q$ from the sampling and for each bit within dit $I$, we find a qubit index $n$ such that $Q=Q_n$ and $I=I_n$.
In \textit{strict} mode, if there is no qubit $n$ we assert that $I$-th bit in $Q$-th dit is \lstinline|0|, which is possible if there are empty places for qubits in the mapping.
Otherwise, in \textit{non-strict} mode, we just ignore bits not in the mapping and set them to \lstinline|0|.
This is equivalent to a partial trace of qudit state over the valid qubit states.
Transpiler steps with qubit-to-qudit mapping, along with qudit samples post-processing (unmapping), represent the main part of the workflow \textit{qubit circuit execution on qudit-based hardware} (Figure~\ref{fig:workflow}). In total, any hardware-agnostic QASM circuit serves as an input to this workflow, with the possibility of using custom qubit-to-quqit mapping. After the workflow execution, we obtain qubit results that can be matched with our input circuit. This approach was introduced in~\cite{Nikolaeva2021} and developed in this paper.

\section{Comparison of Transpilation Approaches}\label{comparison}

To benchmark developed transpiler techniques, in Table~\ref{tab:circuits}, we provide a comparison of output circuits for different commonly used cases and different transpiler options that include optimization, qudit level count $d$, and qubits per qudit parameter $b$.
\textit{The With optimization} column implies all optimization techniques from Section~\ref{transpiler}, whereas \textit{Without optimization} column only decomposes the QASM circuit into \lstinline$IqcCircuit.$
Moreover, we used the Qiskit~\cite{Qiskit} quantum computing framework using \lstinline$qiskit.qasm2.load$ and \lstinline$qiskit.transpile$ functions that convert the QASM circuit to a native gate set for a given target and perform optimizations.
In our case, this was for the trapped ion quantum computer with native gates \rz{}{}, \rp{}{} and \xx{}{}.
However, to our knowledge, the Qiskit framework is not able to produce qudit circuits.
So, it can only be compared with the developed transpiler with parameters $d{=}2$ and $b{=}1$, which is equivalent to the qubit transpilation.

\begin{table*}[ht]
\begin{tabular}{|c|c|c|c|c|c|c|c|}
\hline
\multirow{2}{*}{Name} & \multirow{2}{*}{Regime}
& \multicolumn{3}{c|}{Without optimization} & \multicolumn{3}{c|}{With optimization} \\
& & \rz[]{}{} count & \rp[]{}{} count & \xx[]{}{} count & \rz[]{}{} count & \rp[]{}{} count & \xx[]{}{} count \\
\hline

\multirow{4}{*}{
\shortstack{Bernstein-Vazirani\\ \\identify $101_2$ string}}
    & \textbf{qiskit} & - & - & - & 0 & 24 & 2 \\
    & \textbf{qubit} & 16 & 22 & 2 & 6 & 13 & 2 \\
    & \textbf{qutrit} & 16 & 22 & 2 & 6 & 13 & 2 \\
    & \textbf{ququart} & 25 & 72 & 1 & 5 & 42 & 1 \\
\hline
\multirow{4}{*}{
\shortstack{Bernstein-Vazirani\\ \\identify $10101_2$ string}}
    & \textbf{qiskit} & - & - & - & 0 & 36 & 3 \\
    & \textbf{qubit} & 24 & 32 & 3 & 8 & 19 & 3 \\
    & \textbf{qutrit} & 24 & 32 & 3 & 9 & 19 & 3 \\
    & \textbf{ququart} & 37 & 106 & 2 & 7 & 58 & 2 \\
\hline
\multirow{4}{*}{
\shortstack{Grover\\ \\find $000_2$ string}}
    & \textbf{qiskit} & - & - & - & 4 & 35 & 7 \\
    & \textbf{qubit} & 22 & 48 & 6 & 7 & 13 & 6 \\
    & \textbf{qutrit} & 14 & 42 & 4 & 9 & 18 & 4 \\
    & \textbf{ququart} & 25 & 172 & 4 & 7 & 82 & 4 \\
\hline
\multirow{4}{*}{
\shortstack{Grover\\ \\find $0000_2$ string}}
    & \textbf{qiskit} & - & - & - & 30 & 128 & 40 \\
    & \textbf{qubit} & 93 & 223 & 36 & 37 & 47 & 36 \\
    & \textbf{qutrit} & 29 & 135 & 16 & 18 & 73 & 16 \\
    & \textbf{ququart} & 130 & 704 & 20 & 31 & 353 & 20 \\
\hline
\multirow{4}{*}{
\shortstack{SWAP test\\ \\1 qubit states}}
    & \textbf{qiskit} & - & - & - & 3 & 24 & 7 \\
    & \textbf{qubit} & 22 & 42 & 7 & 8 & 11 & 7 \\
    & \textbf{qutrit} & 14 & 36 & 5 & 4 & 15 & 5 \\
    & \textbf{ququart} & 22 & 132 & 4 & 1 & 59 & 4 \\
\hline
\multirow{4}{*}{
\shortstack{SWAP test\\ \\2 qubits states}}
    & \textbf{qiskit} & - & - & - & 6 & 47 & 14 \\
    & \textbf{qubit} & 44 & 82 & 14 & 15 & 19 & 14 \\
    & \textbf{qutrit} & 28 & 70 & 10 & 7 & 28 & 10 \\
    & \textbf{ququart} & 44 & 256 & 8 & 1 & 114 & 8 \\
\hline
\end{tabular}
\caption{
Comparison of transpilation for different regimes.
We produce the circuits for each case and permute qubits manually in order to obtain  a minimum \xx{}{} count in the \textbf{ququart} regime.
This permutation does not affect the gate count in \textbf{qiskit}, \textbf{qubit} and \textbf{qutrit} regimes.
}
\label{tab:circuits}
\end{table*}

We choose the following list of circuits for a comparison of different transpilation regimes:
\begin{itemize}
    \item \textit{Bernstein--Vazirani algorithm}: \begin{itemize}
        \item The identification of $101_2$ binary string on 3 qubits and 1 ancillary qubit;
        \item The identification of $10101_2$ binary string on 5 qubits and 1 ancillary qubit.
    \end{itemize}
    \item \textit{Grover algorithm}: \begin{itemize}
        \item Finding $000_2$ binary string on 3 qubits and 1 ancillary qubit;
        \item Finding $0000_2$ binary string on 4 qubits and 1 ancillary qubit.
    \end{itemize}
    \item \textit{Swap test}: \begin{itemize}
        \item Orthogonal qubit states on 3 qubits;
        \item Orthogonal 2-qubits states on 5 qubits.
    \end{itemize}
\end{itemize}
These input qubit circuits have a representation in QASM format, which is the same for every regime:
\begin{itemize}
    \item \textbf{qiskit}: \lstinline$qiskit.qasm2.load$, \lstinline$qiskit.transpile$ with trapped-ion basic gates;
    \item \textbf{qubit}: level count $d{=}2$, qubits per qudit $b{=}1$;
    \item \textbf{qutrit}: level count $d{=}3$, qubits per qudit $b{=}1$;
    \item \textbf{ququart}: level count $d{=}4$, qubits per qudit $b{=}2$.
\end{itemize}

We note that the \textbf{qiskit} regime is only shown as optimized since \lstinline$qiskit.transpile$ always optimizes the input circuit.

Firstly, the effect of circuit optimization can be clearly observed.
For every circuit from the list, an optimization pass reduces the number of \rp{}{} gates at least by 41\%.
Furthermore, some cases have a reduction of up to 79\% in gate amount in the case of Grover's algorithm for \textbf{qubit} regime.
Notably, the whole circuit can be optimized if the optimizer decides that it is equivalent to the \id{} operation. In these experiments, we did not use barriers in the circuits.

Secondly, our transpiler shows better output (in terms of the amount of \rp{}{} gates)  than the \lstinline$qiskit.transpile$ function.
However, this result is obtained by the increased number of \rz{}{} gates in an output circuit.
However, regarding the fact that \rz{}{} gates are implemented virtually on qudit quantum computer, our optimization results can be considered more effective.

Thirdly, \textbf{qutrit} and \textbf{ququart} regimes can also optimize the amount of \xx{}{} gates.
This result has more impact since two-qudit gates are more subjected to errors than single-qudit gates.
This effect takes place in circuits with a large number of qubits.
However, the single-qudit gate amount for these regimes is much greater than the \textbf{qubit} amount.
This is due to our optimization workflow that only optimizes gates on the same levels, but the circuit can contain gates on all pairs of levels.
One can consider a simplistic approach to choosing a preferable regime for running a circuit. Given estimates of single-qubit and two-qubit gate errors ($e_{\rm 1b}$ and $e_{\rm 2b}$), as well as single-qudit and two-qudit gate errors ($e_{\rm 1d}$ and $e_{\rm 2d}$), the resulting errors in a qubit-based and qudit-based circuit can be approximately upper-bounded as
\begin{equation}
    \begin{aligned}
        E_{\rm b}&=e_{1\rm b} N_{1\rm b}+e_{2\rm b} N_{2\rm b},\\
        E_{\rm d}&=e_{1\rm d} N_{1\rm d}+e_{2\rm d} N_{2\rm d},\\
    \end{aligned}
\end{equation}
where $N_{1{\rm b(d)}}$ and $N_{2{\rm b(d)}}$ are the numbers of single-qubit (qudit) and two-qubit (qudit) gates correspondingly.
Comparing $E_{\rm b}$ and $E_{\rm b}$ can help one to choose the best way to run the circuit on the available hardware. We note that $N_{\rm 1b(d)} $ can be calculated without including virtual phase gates.
We also note that, due to the mentioned fact that the optimization workflow optimizes gates at the same level, but a circuit may contain gates between all pairs of levels, the idea of \textit{whole qudit optimization} arises. This can decompose any sequence of operations or a general single-qudit unitary matrix into the most optimal sequence in terms of the number of \rp{}{} gate amount, as was conducted in~\cite{Younis2023}.

It is worth noting that the performance of the \textbf{ququart} regime is significantly influenced by the qubit-to-qudit mapping. Specifically, the way in which qubits are distributed among qudits has a significant impact on the resulting number of entangling gates.
Within a straightforward mapping, for example, three qubits that are affected by the Fredkin \gate{CSWAP}{}{}{} gate in the \textit{Swap test (2 qubit states)} algorithm are placed in separate ququarts, resulting in 16 $\xx{}{}$ gates in the qudit circuit. However, with a more carefully designed mapping, it is possible to store two of these qubits in a single ququart, reducing the number of $\xx{}{}$ gates to 8.
This example demonstrates that different mappings can lead to better or worse results than the \textbf{qubit} regime. Therefore, finding an optimal mapping for a particular circuit is crucial when running a qubit circuit on a qudit platform.

\section{Conclusions}\label{conclusion}

In this paper, we provided the general workflow for converting qubit QASM circuits into qudit circuits and compared transpilation for different regimes.
With different values for parameters $d$ (dimension of qudit) and $b$ (qubit count in a single qudit), the transpiler uses different decomposition approaches and produces a qudit circuit, which can be executed on a quantum computer with qudit dimension $\geq d$.
Parameters $d$ and $b$ should be chosen reasonably for a given problem.
The first example is that circuits with a  high arity of \mcx{}/\mcz{} gates benefit from the \textbf{qutrit} regime since it decomposes a multiqubit gate into $2N-3$ two-qudit operations, whereas \textbf{qubit} decomposition has at least \On{N} two-qudit gates.
The second example is circuits with distinct pairs of qubits with a high weight of interconnection (amount of two-qubit gates) within a pair and low interconnection between pairs.
Using the \textbf{ququart} regime and an appropriate qubit-to-qudit mapping can greatly reduce the amount of two-qudit gates compared to the \textbf{qubit} regime, although optimal mapping could be performed automatically (locally for each \mcz{} and globally for a whole circuit).
Since naive mapping optimization via finding the best qubit permutation requires \On{N!}, we leave the nearly optimal mapping finding feature to future versions of the transpiler.

We also consider qudit circuit optimization reduces qudit gate amount in the resulting circuit.
For different regimes, the reducing factor is in the range of 40\% to 80\%.
Optimization routine accounts allowed transitions for \rp{}{} and \xx{}{} gates for a given target qudit system.
However, current optimization can be further improved to produce the most optimal output.
This is the main focus for future work.

\section*{Acknowledgements}
The work of D.A.D. and A.S.N. was supported by RSF Grant No.~24-71-00084 (developing data post-processing methods for qudit-based architectures).
E.O.K. and A.K.F. acknowledge support from the Priority 2030 program at the NIST ``MISIS'' under the project K1-2022-027.

The authors declare no conflicts of interest.

\bibliography{bibliography-qudits.bib}

\newpage
\appendix

\section{QASM Format Instructions}\label{app:qasm-format}

Open Quantum Assembly format~\cite{OpenQASM} is commonly used for quantum circuit description.
It allows us to represent a circuit as a sequence of instructions.

\textit{Quantum and classical register declarations}.
These instructions define the use of qubits and classical bits in the following circuit:
Users can specify them by the register name and register size:
\begin{itemize}
    \item Quantum register \lstinline|qreg q[N];|
    \item Classical register \lstinline|creg c[M].|
\end{itemize}

\textit{Quantum operations' definitions and declarations}.
These statements describe which gates can be used in the circuit and define their parameters.
\lstinline|opaque| operation states that a gate is already presented on the target device, and \lstinline|gate| operation defines gate decomposition in terms of previously defined gates and opaque operations.
Typically, all useful quantum gates are defined in \lstinline|"qelib1.inc"|:
\begin{itemize}
    \item Gate definition \lstinline|gate U(a, b, c) q { ... }|
    \item Gate declaration \lstinline|opaque U(a, b, c) q;|
\end{itemize}

\textit{Apply gate instruction}.
With predefined quantum registers and gates, users are able to construct a circuit by applying a gate to a qubit or a set of qubits:
\begin{itemize}
    \item \lstinline|X q[0]; U(0.0, 1.0, 2.0) q[2];|
    \item \lstinline|CX q[0], q[1];|
\end{itemize}

\textit{Barrier statement}.
Special \lstinline|barrier| statement does nothing in a quantum circuit and prevents an optimization by gate merging:
\begin{itemize}
    \item \lstinline|barrier q[0], q[2];|
\end{itemize}

\textit{Reset statement}.
Collapses qubit state and sets it to $\ket{0}$:
\begin{itemize}
    \item \lstinline|reset q[0];|
\end{itemize}

\textit{Measure statement}.
Performs a measurement of qubit and places its result (\lstinline|0| or \lstinline|1|) in a classical bit.
At the end of experiments, users should measure qubits and store results in classical bits since QASM semantics states that quantum computer should produce samples according to measured classical bits:
\begin{itemize}
    \item \lstinline|measure q[0] -> c[0];|
\end{itemize}

\textit{Conditional statement}.
Determines whether the inner statement runs according to value in a classical bit:
\begin{itemize}
    \item \lstinline|if (c[0] == 1) X q[0];|
    \item \lstinline|if (c[0] == 1) reset q[0];|
\end{itemize}

\section{Ion Quantum Computer Circuit Description Format}\label{app:iqc-json-format}

Qudit circuits for an ion quantum computer can be written as a JSON file containing an array of objects:
\begin{equation*} \label{iqc}
    \verb$IqcJsonFormat = IqcCircuit[]$
\end{equation*}
Each object \lstinline$IqcCircuit$ represents a single circuit to execute; hence, it is allowed to describe and execute several circuits in a single file.
Circuit contains information, which could be used to execute qudit circuit:
\begin{equation*} \label{iqc1}
\begin{aligned}
    & \verb$IqcCircuit = {$ \\
    & \verb$  "repetitions": integer,$ \\
    & \verb$  "levels": integer,$ \\
    & \verb$  "sequence": IqcOperation[]$ \\
    & \verb|}|
\end{aligned}
\end{equation*}
where \lstinline$"repetitions"$ parameter denotes the number of shots to collect circuit's statistics, \lstinline$"levels"$ denotes qudit level count $d$ and \lstinline$"sequence"$ represents the sequence of qudit operations itself.

Operation \lstinline$IqcOperation$ is equivalent to one of the qudit gates given by Equation~(\ref{eq:ion}) with the given angles and levels parameters:
\begin{equation*} \label{iqc2}
\begin{aligned}
    & \verb$IqcOperation =$\\
    & \verb$  IqcPhGate$ \\
    & \verb$  | IqcRGate$ \\
    & \verb$  | IqcXXGate$
\end{aligned}
\end{equation*}
\begin{equation*} \label{iqc3}
\begin{aligned}
    & \verb$IqcPhGate = {$ \\
    & \verb$  "type": "Rz",$ \\
    & \verb$  "angle": float,$ \\
    & \verb$  "upper_state": integer,$ \\
    & \verb$  "qudit": integer$ \\
    & \verb|}|
\end{aligned}
\end{equation*}
\begin{equation*} \label{iqc4}
\begin{aligned}
    & \verb$IqcRGate = {$ \\
    & \verb$  "type": "Rphi",$ \\
    & \verb$  "angle": float,$ \\
    & \verb$  "axis": float,$ \\
    & \verb$  "upper_state": integer,$ \\
    & \verb$  "qudit": integer$ \\
    & \verb|}|
\end{aligned}
\end{equation*}
\begin{equation*} \label{iqc5}
\begin{aligned}
    & \verb$IqcXXGate = {$ \\
    & \verb$  "type": "XX",$ \\
    & \verb$  "angle": float,$ \\
    & \verb$  "upper_state": integer,$ \\
    & \verb$  "qudits": integer[]$ \\
    & \verb|}|
\end{aligned}
\end{equation*}
where \lstinline$"angle"$ and \lstinline$"axis"$ parameters represent $\theta$ and $\phi$ from Equation~(\ref{eq:ion}), respectively; divided by $\pi$, \lstinline$"upper_state"$ stands for level $i$ with the exception of \lstinline$IqcXXGate$, where  \lstinline$"upper_state": 1$ is the only supported level.
\lstinline$"qudit"$ and \lstinline$"qudits"$ are the numbers of qudqudits on which the given operation is acting.

\section{Transpiler's Runtime Description: \textit{qelib1.inc} and \textit{matcher.script}}\label{app:runtimes}

\textit{qelib1.inc} is the standard included file from the QASM specification.
It has the purpose of storing all definitions of gates (\lstinline$opaque$ or \lstinline$gate$ instructions), which can be used to construct quantum circuits.
There is no regregulation on how to define a native gate set in QASM, yet we use the following notations:
\begin{itemize}
    \item \lstinline$opaque$---Declares gate without definition.
    We consider these declarations as native gates of the given device.
    \item \lstinline$gate$---Defines gate in terms of previously declared gates.
    This can be any gate (\gate{U}{}{}{}, \gate{CX}{}{}{}, Toffoli, Fredking).
\end{itemize}
Transpile uses this notation in the first phase of the qubit transpilation step. Any gate call in the QASM file that it sees is either \textit{opaque} or \lstinline$gate$.
Each \lstinline$gate$ is decomposed into the sequence of gate calls.
This process runs until there are only \lstinline$opaque$ gates (also known as native gates) in the circuit.

Our \textit{qelib1.inc} for \textit{trapped ion intermediate representation} is depicted in Figure~\ref{fig:qelib1-code}.

\begin{figure}[b]
\centering
\begin{lstlisting}
opaque rz(the) q0;
opaque r(the, phi) q0;
opaque cz q0, q1;
opaque ccz q0, q1, q2;
opaque cccz q0, q1, q2, q3;
...
gate x q0 { r(pi, 0) q0; }
gate z q0 { rz(pi) q0; }
gate h q0 {
  r(pi/2, -pi/2) q0;
  rz(pi) q0;
}

gate rx(the) q0 {
  r(the, 0) q0;
}
gate ry(the) q0 {
  r(the, pi/2) q0;
}
gate u3(the, phi, lam) q0 {
  rz(lam) q0;
  ry(the) q0;
  rz(phi) q0;
}
...
gate cx q0, q1 {
  h q1;
  cz q0, q1;
  h q1;
}
gate ccx q0, q1, q2 {
  h q2;
  ccz q0, q1, q2;
  h q2;
}
gate cswap q0, q1, q2 {
  cx q2, q1;
  ccx q0, q1, q2;
  cx q2, q1;
}
...
\end{lstlisting}
\caption{Examples of gate definitions from \textit{trapped ion intermediate representation} runtime.
Native gates for this runtime are labeled with \textbf{opaque} keyword.}
\label{fig:qelib1-code}
\end{figure}

The \textit{matcher.script} file is the extension to a target device description that is used in the optimization phase of qubit transpilation as the list of optimization rules.
Each rule has one of the following properties:
\begin{itemize}
    \item Reducing sequence length:\\
        $\rz[\theta_1]{}{}\circ\rz[\theta_2]{}{}\rightarrow\rz[\theta_1 + \theta_2]{}{}$.
    \item Replacing complex gates with simpler equivalent sequences:\\     $\rp[\theta_1, \phi_1]{}{}\circ\rp[\theta_2, \phi_2]{}{}\rightarrow\rp[\theta', \phi']{}{}\circ\rz[\theta'']{}{}$.
    \item Replacing gate parameters according to symmetries:\\
        $\rp[\theta + 4\pi, \phi]{}{}\rightarrow\rp[\theta, \phi]{}{}\quad\rp[0, \phi]{}{}\rightarrow\id$.
    \item Preserving some ordering in a sequence (in this case, the transpiler will try to move \rz{}{} to the end of the circuit):\\
        $\rz[\varphi]{}{}\circ\rp[\theta, \phi]{}{}\rightarrow\rp[\theta, \phi - \varphi]{}{}\circ\rz[\varphi]{}{}$\\
        but not \\
        $\rp[\theta, \phi]{}{}\circ\rz[\varphi]{}{}\rightarrow\rz[\varphi]{}{}\circ\rp[\theta, \phi + \varphi]{}{}$.
\end{itemize}

To implement these rules in \textit{matcher.script}, we defined a special description language, similar to C and QASM.
Examples of the rules are depicted in Figure~\ref{fig:matcher-code}.

The transpiler operates with \textit{the circuit pattern}, which is a sequence of gates acting on the same set of qubits.
In \textit{matcher.script}, it is represented as a composition of gates using the \textit{bullet} ``\lstinline$.$'' operator.
For example, ``\lstinline$rz(a) x . cz x, y$'', where \lstinline$rz$ and \lstinline$cz$ are native gates from selected runtime, \lstinline$x$ and \lstinline$y$ are qubit patterns and \lstinline$a$ is a parameter pattern.
The matcher goes through each subsequence of gates in the original circuit and tries to find rules to apply and acquire actual values for qubit and parameter patterns.
With these values, the matcher runs a code block from the source code and obtains a new sequence of gates via the \lstinline$return$ instruction.
The returned value is also a composition of gates using \textit{bullet}.
However, all the qubits and parameters already have concrete values.

\begin{figure}[t]
\centering
\begin{lstlisting}[language=C]
// Reducing sequence length:
rz(a0) . rz(a1) => {
  return rz(a0 + a1);
}
...
// Replacing gate parameters
// according to symmetries:
rz(a) => {
  a_2 = a / 2;
  s = sin(a_2);
  if s == 0 {
    return id;
  } else if a_2 > pi || a_2 < -pi {
    return rz(2 * atan2(s, cos(a_2)));
  }
}
...
// Preserving some ordering in a sequence:
rz(a) . r(b, c) => {
  return r(b, c - a) . rz(a);
}
rz(a) x . cz x, => {
  return cz x,y . rz(a) x;
}
rz(a) y . cz x,y => {
  return cz x,y . rz(a) y;
}
\end{lstlisting}
\caption{Examples of optimization rules from \textit{trapped ion intermediate representation} runtime.}
\label{fig:matcher-code}
\end{figure}

\end{document}